\documentclass[12pt,preprint]{aastex}
\usepackage{float,epsfig} 
\usepackage{graphicx}
\usepackage{epstopdf}

\def\gtrsim{\mathrel{\hbox{\rlap{\hbox{\lower4pt\hbox{$\sim$}}}\hbox{$>$}}}}
\def\lesssim{\mathrel{\hbox{\rlap{\hbox{\lower4pt\hbox{$\sim$}}}\hbox{$<$}}}}

\newcommand{\beq}{\begin{equation}}
\newcommand{\eeq}{\end{equation}}

\begin{document}

\title{Black hole quasinormal mode spectroscopy with LISA}
\author{Manish M.~Jadhav\altaffilmark{1} and 
Lior M.~Burko\altaffilmark{1,2}}
\altaffiltext{1}{Department of Physics, University of Alabama in Huntsville, Huntsville, AL 35899, USA}
\altaffiltext{2}{Center for Space Plasma and Aeronomic Research, University of Alabama in Huntsville, Huntsville, AL 35899, 
USA}

\begin{abstract}
The signal--to--noise ratio (SNR)  for black hole quasinormal mode sources of low--frequency gravitational waves is estimated using a Monte Carlo approach that replaces the all--sky average approximation. We consider an eleven dimensional parameter space that includes both source and detector parameters. We find that in the black--hole mass range $M\sim 4$--$7\times 10^6M_{\odot}$ the SNR is significantly higher than the SNR for the all--sky average case, as a result of the variation of the spin parameter of the sources. This increased SNR may translate to a higher event rate for the Laser Interferometer Space Antenna (LISA). We also study the directional dependence of the SNR, show at which directions in the sky LISA will have greater response, and identify the LISA blind spots.
\end{abstract}

\keywords{gravitational waves --- black hole physics}

\section{Introduction}
An important source of gravitational waves in the frequency range $10^{-4}$--$10^{-1}$Hz, which is the frequency band for the planned space-bourne gravitational-wave detector Laser Interferometer Space Antenna (LISA), is the inspiral and merger of two supermassive black holes. The emitted gravitational waveform includes three major parts, corresponding to the inspiral, merger, and ringdown of the system. The last stage, the ringdown, is the settling down of the resulting black hole to quiescence as required by the ``no-hair'' theorem, and is characterized by (an incomplete) set of complex frequencies that depend only on the macroscopic parameters of the black hole, namely its mass $M$ and spin angular momentum $J$ (Nollert 1999). The computation of the complex frequency is done using black hole perturbation theory (Chandrasekhar 1983), and its detection would be a smoking gun for the source being a black hole, and an accurate means for determining its parameters (Berti et al.~2006). 

Naively one would think that black hole quasinormal modes are not a strong source of gravitational waves for LISA. Indeed, the typical (real part of the) frequency for a Schwarzschild black hole (for quadrupole radiation $\ell=2$ and azimuthal mode $m=2$ at the longest damping time $n=0$) is $f_{\ell=2,m=2,n=0}\sim 1.2\times 10^4 (M_{\odot} / M)$Hz whereas the e-folding time for the exponential decay of the amplitude (associated with the imaginary part of the complex frequency) is  $\tau_{\ell=2,m=2,n=0}\sim 55 (M / M_{\odot})\mu{\rm s}$. While for a stellar black hole this frequency is several orders of magnitude too high to be detected by LISA, the frequency for supermassive black holes falls inside LISA's good sensitivity frequency band. The quality factor $Q$ associated with the oscillator is $Q:=\pi f\tau$, or for a Schwarzschild black hole $Q\approx 2$, i.e., within two oscillations the amplitude of the waves is suppressed by a factor of $e^{2\pi}\sim 535$. (While for rotating black holes the quality factor is higher, it is still low unless the black hole's spin approaches extremality.  Indeed, the quality factor can be approximated by $Q_{220}=0.7000+1.4187(1-j)^{-0.4990}\pm 0.88\%$ (Berti et al.~2006). A different parameterization of $Q_{220}$, which however agrees with this one to within $1\%$ was first given by Echeverria (1998).) 
Therefore, a black hole is expected to be a very poor resonator, and the waves difficult to detect. Regardless of these naive expectations, Flanagan \& Hughes (1998) showed that the signal-to-noise ratio (henceforth SNR) from the ringdown stage could be comparable to that of the inspiral stage, so that waves emitted during the ringdown  stage may be an important source for LISA. 

The signal to be detected by LISA comes from sources that span a large parameter space, including both source parameters and detector parameters. To find the SNR from a large sample of ringdown sources as a function of the black hole's mass one needs to  include our ignorance of the particular parameters. Specifically, we do not know the parameters $\theta,\phi$ that are the spherical polar angles that refer to the direction to the source, and $\psi$ which is the angle between the waves polarization plane and the plane $\phi=0$ that determine the detector pattern functions $F^{+,\times}(\theta,\phi,\psi)$ for the two gravitational wave polarization states. Specifically, 
$$F^{+}(\theta,\phi,\psi)=\frac{1}{2}(1+\,\cos^2\theta)\,\cos 2\phi\,\cos 2\psi-\,\cos\theta\,\sin 2\phi\,\sin 2\psi$$
$$F^{\times}(\theta,\phi,\psi)=\frac{1}{2}(1+\,\cos^2\theta)\,\cos 2\phi\,\cos 2\psi+\,\cos\theta\,\sin 2\phi\,\sin 2\psi\, .$$
We also are ignorant of the source's parameters $\iota,\beta$ that are the spherical polar angles in the frame of the source, defining the orientation of the source in its local frame (i.e., determining the orientation of the black hole's spin axis). Other unknown parameters are the black hole's dimensionless specific spin angular momentum $j:=cJ/(GM^2)$, the phases of the two polarization states $\phi^{+,\times}_{\ell,m,n}$, the total radiation efficiency $\epsilon_{\rm rd}:=M^{-1}c^{-2}\int_0^{\infty}(\,dE/\,df)\,df$ (i.e., the fraction of $Mc^2$ being emitted in gravitational waves) and $\epsilon^+$, the fraction thereof in the $+$ polarization state, in addition to the distance to the source, which we take to be the luminosity distance $D_L$. Altogether, we have an eleven dimensional parameter space (3 dimensions for the detector parameters, and 8 for the source's) (Thorne 1987).

The currently available method is to make the ignorance of the parameters manifest by averaging over some of the parameters (the detector parameters, i.e., $\theta$, $\phi$, and $\psi$, in addition to the the two source parameters $\iota$ and $\beta$) and fixing other parameters (the source parameters $j$, $\phi^{+,\times}_{\ell,m,n}$, $\epsilon_{\rm rd}$, $\epsilon^{+}$, $D_L$). Consider the question of what is the SNR for ringdown sources of a particular mass. This question is important for detection purposes or for estimation of the population statistics.  (This question is different from the question of what is the SNR for a ringdown source with particular parameters, which we address in what follows.) 
Specifically, as one black hole  source may be in one direction in the sky with a certain value for the specific spin (and similarly for the other parameters), and another black hole source may be in a different direction with a different value of the specific spin (and the other parameters), one may assume that all black hole sources have average (or fixed) values for all parameters (notice that Berti et al.~(2006) did not average over the spin angular momentum $j$, but rather fixed it).   Hereafter, we refer to this approach as the ``all--sky average" (because it involves averaging over the detector parameters, although it involves also parameter fixing and not just averaging).  In this approach one would argue that a good estimate for the total SNR for ringdown sources in a given mass interval would be found by taking the spin of all such sources to be the average value, say $j=0.5$ if the distribution function is taken to be uniform. 
The all--sky average approach was used recently by Berti et al.~(2006) to find the SNR. The all--sky average approach clearly is useful as an approximation method to find the SNR, and as we show below does it to very high accuracy. Notice, that our ignorance of the parameters to be used is of two kinds: first, there are observational unknowns, that will become known after precise observations are made. Second, there are theoretical unknowns, that would become better known (including their distribution functions) after better theoretical modeling and understanding of the sources' physics is obtained. We do not distinguish here between the two types of unknowns. 

An alternative approach is to integrate over a finite number of sources, and use Monte Carlo integration to reflect the ignorance of the parameters. That is, one uses random values of the parameters for each source, spanning the parameter space, and then averages over a large number of such sources. The range of the parameters and the distribution functions for the random sample then reflect the physics of the source and the gravitational--wave generation. In practice, we take here for simplicity all distribution functions to be uniform, but this assumption is easy to relax if evidence is presented for non-uniform distributions. Specifically, we take the sources to be homogeneous and isotropic (so that the uniformity of the distribution function is in $D_L^2$ and $\cos\theta$), and the sources' spin axes to be isotropic (i.e., the distribution function is homogeneous in $\cos\iota$). Recently, our understanding of gravitational--wave sources has improved dramatically. We now know that the radiation efficiency for binary inspiral, at least for the variables simulated using numerical relativity, is $\epsilon_{\rm rd}\sim 3\%$ with circular polarization. However, it was argued that $\epsilon_{\rm rd}\gtrsim 7\%$ for some cases (Berti et al.~2007), and for head--on collision it is $\epsilon_{\rm rd}\sim 0.1\%$ with $+$ polarization (Berti et al.~2006). 
More recent numerical relativity simulations for the merger of comparable mass black holes have found a relationship of the gravitational--wave radiation efficiency to the spin of the initial black holes, and a relationship of the spin of the final black hole to the spin of the initial black holes (Campanelli et al.~2007). For the specific case of initially aligned or anti-aligned black hole spins, the relationship of the radiation efficiency in gravitational waves to the final black hole's spin $j$ is given by 
\begin{equation}\label{rad}
\epsilon_{\rm rd}(j) = 0.994-0.438\,\sqrt{4.751-3.74\,j}-0.4545\,j\, ,
\end{equation}
as obtained from a quadratic fit of the numerical relativity results of Campanelli et al.~(2007). The final black hole's spin was found to be in the range $0.35\lesssim j\lesssim 0.95$, which implies radiation efficiency in the range $2.2\%\lesssim 
\epsilon_{\rm rd}\lesssim 8.3\%$. The dependence $\epsilon_{\rm rd}=\epsilon_{\rm rd}(j)$ reduces our parameter space from 11 to 10 independent dimensions. 
In what follows, we specialize our discussion to the case of quasi--normal modes from black hole sources resulting from the merger of black hole binaries with initially aligned or anti-aligned spins. This specialization allows us to use actual numerical relativity values for the radiation efficiency in gravitational waves. Our method can be applied also for black holes resulting from less restrictive initial spin configurations when corresponding numerical relativity results become available.

One may naively expect that in the limit of very many sources, the Monte Carlo results reduce to the all--sky average results. However, they do not. The large number limit of the Monte Carlo integration approaches the all--sky average only if the parameters of the SNR are independent. When considering only Schwarzschild black holes, or even all black holes with a fixed  value for the specific spin angular momentum, indeed all the parameters are independent, i.e., they live in different spaces. However, when one considers that for any given value for the black hole mass $M$ the specific spin spans the range $0\leqslant j\leqslant 1$ (or, more specifically for the case considered here in detail, $0.35\lesssim j\lesssim 0.95$), these parameters are no longer independent. 

Specifically, the SNR squared is given by (Berti et al.~2006)
\begin{eqnarray}
\label{snr}
\rho^2 &=&  2\left(\frac{GM}{rc^2}\right)^2\int_0^\infty\frac{df}{S_h(f)}\Bigg\{(b_+^2 + b_{-}^2)\left[{A^+_{\ell mn}}^2{F^+}^2  \right. \nonumber \\
&+& \left. {A^\times_{\ell mn}}^2{F^\times}^2 - 2{A^+_{\ell mn}}{A^\times_{\ell mn}}{F^+}{F^\times}\sin(\phi^+_{\ell mn} - \phi^\times_{\ell mn})\right]\times\left| S_{\ell  mn} \right|^2   \nonumber \\
&+&  2b_+b_{-}\left\{\Re\left[\left({A^+_{\ell mn}}^2{F^+}^2e^{2i\phi^+_{\ell mn}} - {A^\times_{\ell mn}}^2{F^\times}^2e^{2i\phi^\times_{lmn}}\right){S_{\ell mn}}^2\right]\right. \nonumber \\
&+& \left. 2{A^+_{\ell mn}}{A^\times_{\ell mn}}{F^+}{F^\times}\Im\left[e^{i(\phi^+_{\ell mn} + \phi^\times_{\ell mn})}{S_{\ell mn}}^2\right] \right\}\Bigg\}\, .  
\end{eqnarray}
Here, $b_{\pm}=(1/\tau_{\ell mn})/[(1/\tau_{\ell mn})^2+4\pi^2(f\pm f_{\ell mn})^2]$ is the Fourier transform of the time domain single frequency quasi-normal signal that following Berti et al.~(2006) we approximate using the delta--function (in frequency) approximation, and $S_h(f)$ is the LISA noise spectral density. In practice we calculate $S_h(f)$ using the approximation given by 
Finn \& Thorne (2000) and by Larson et al.~(2000), which agrees well, except for oscillations at the high--frequency end, with the LISA Sensitivity Curve Generator results\footnote{The LISA Sensitivity Curve Generator (SCG) is available online at this URL: {\tt http://www.srl.caltech.edu/$\sim$shane/sensitivity/MakeCurve.html}}. The amplitudes in the two polarization states are denoted by $A^{\pm}_{\ell mn}$, and $S_{\ell mn}(\iota,\beta)$ are the spin--weighted spheroidal harmonics of spin weight 2, which depend on $j$ through the overtone index $n$. Also the amplitudes $A^{\pm}_{\ell mn}$ are functions of $j$, as 
\begin{eqnarray}
\label{amp}
A^{+}_{\ell mn}  =  \sqrt{\frac{8c^3\,\epsilon_{\rm rd}^{+}(1 + 4Q_{\ell mn}^2)}{GMQ_{\ell mn}f_{\ell mn}(1 + 2Q_{\ell mn}^2)}}\;\;\;\;\; , \;\;\;\;\;
A^{\times}_{\ell mn}  =  \sqrt{\frac{4c^3\,\epsilon_{\rm rd}^{\times}(1 + 4Q_{\ell mn}^2)}{GMQ_{\ell mn}^3f_{\ell mn}}}
\end{eqnarray}
and all three variables $f_{\ell mn}$, $Q_{\ell mn}$, and $\epsilon_{\rm rd}$ are functions of $j$. Clearly, $ \langle A_{\ell mn}(j)\,S_{\ell mn}(j) \rangle  \neq  \langle A_{\ell mn}(j) \rangle\times \langle S_{\ell mn}(j) \rangle$ when the averaging is done over $j$. We may not therefore expect the qualitative dependence of $\rho^2$ as a function of the mass $M$ to remain unchanged when the all--sky averaging assumption is relaxed. In this Paper we study this question, and find the SNR when the sources are obtained from a random sample. That is, we do not assume following Berti et al.~(2006), that all black hole quasi-normal mode sources have the average or fixed values for all parameters, and we let these parameters take random values, as one would expect from an actual sample of sources for LISA. 
The all--sky average approximation of Berti et al.~(2006) assumed that ${F^{+,\times}}^2=\langle  {F^{+,\times}}^2\rangle=\frac{1}{5}$, 
${F^{+}F^{\times}}=\langle  {F^{+}F^{\times}}\rangle=0$, and that $S_{\ell mn}^2=S_{\ell mn}^*\,S_{\ell mn}=\langle  S_{\ell mn}^*\,S_{\ell mn} \rangle=\frac{1}{4\pi}$, where an asterisk denotes complex conjugation.  It was further assumed following Flanagan \& Hughes (1998) that $\phi^+_{\ell mn}=\phi^\times_{\ell mn}=0$ and that the amplitudes $A^+_{\ell mn}=A^\times_{\ell mn}=A_{\ell mn}$. The specific spin angular momentum $j$ and the radiation efficiency $\epsilon_{\rm rd}$ were fixed. 
To compare our Monte Carlo results results with the all--sky average approach, we set the spin angular momentum in the all--sky average case to the value that gives the average radiation efficiency. Specifically, we find $j$ that would solve 
$\epsilon_{\rm rd}(j)=\langle\epsilon_{\rm rd}(j)\rangle=\,\int_{j_{\rm min}}^{j_{\rm max}}\epsilon_{\rm rd}(j')\,dj'/(j_{\rm max}-j_{\rm min})$. Over the range of $j$ for which $\epsilon_{\rm rd}(j)$ is valid, we find in practice that $\langle\epsilon_{\rm rd}(j)\rangle=0.03824$ and the corresponding spin angular momentum is $j=0.71882$. 

Our Monte Carlo approach is a simple application of Monte Carlo integration, as we are using uniform distribution functions for our random variables (see below for justification). The (pseudo-)random number generator is based on the Mersenne Twister algorithm (Matsumoto \& Nishimura 1998). While not considered secure for cryptography applications, this generator is certainly reliable enough for our purposes. Specifically, we sample points uniformly from the integration region to estimate the integral and its error. Namely, we calculate the SNR  by estimating the integral in (\ref{snr}) by summing the integrand over the randomly chosen sample, dividing by the volume of the parameter space and by the number of iterations. The estimated error $\sigma$ in the Monte Carlo integration is done by calculating the variance, according to $\sigma^2=(V/N)^2\,\sum_{i=1}^{N}[\rho(x_i)-\langle \rho\rangle]^2$, where $V$ is the volume of the parameter space and $N$ is the number of elements in the sample. Here $x_i$ denotes collectively the eleven parameters for the SNR for the $i^{\rm th}$ member of the sample. The reason why this very simple approach to Monte Carlo integration is sufficient for our needs is that we choose the sources to be uniformly distributed across the parameter space. Specifically, we take the sources to be distributed homogeneously and isotropically, which seems to be a very reasonable assumption due to the cosmological nature of the sources. We also take the spin axis of the black hole members of the sample to be distributed isotropically. These assumptions mean that we take $D_L^2$, $\,\cos\theta$, $\,\cos\iota$, $\phi$ and $\beta$ to be uniformly distributed. We have little motivation to prefer a non-uniform distribution function for the other parameters. Specifically, we currently do not have enough numerical relativity results to determine a realistic distribution function for the parameters $\psi, \phi^{+,\times}$ or $\epsilon^+$ in addition to the population statistics of $j$. We therefore take these parameters to be distributed uniformly in their respective parameter spaces. The radiation efficiency $\epsilon_{\rm rd}(j)$ is determined according to Eq.~(\ref{rad}). 

Our approach allows us also to study sources with particular parameters instead of the all--sky averaged counterparts. Specifically, we consider the directional variations of fixed sources, and show in which directions in the sky LISA would be more sensitive, including finding LISA blind spots.

In all our simulations we take a standard zero--curvature cosmological model with the 5--year WMAP values 
$\Omega_M=0.279$, $\Omega_{\Lambda}=0.721$, and 
$H_0=70.1\;{\rm km}\,{\rm s}^{-1}\,{\rm Mpc}^{-1}$ (Hinshaw et al.~2009), and the luminosity distance -- redshift relation to be given by (Carroll et al.~1992)
$$D_L(z)=(1+z)\frac{c}{H_0}\int_0^z\frac{\,dz'}{\sqrt{(1+z')^2(1+\Omega_M z')-z'(2+z')\Omega_{\Lambda}}}\, .$$

\section{Testing the Monte Carlo integration in the Schwarzschild and Kerr cases}
\label{test}

\begin{figure}[h]
 \includegraphics[width=3.4in]{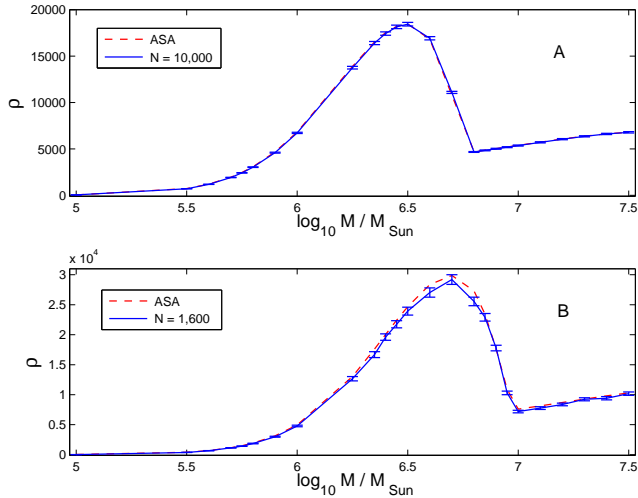} 
\caption{Upper panel (A): The SNR for the all--sky average (results analogous to those of Berti et al.~(2006)) (dashed curve) and the Monte Carlo results with $N=10\, 000$ Schwarzschild sources (solid curves and error bars) at luminosity distance of 3Gpc (redshift of $z=0.522$). Radiation efficiency for the all--sky average case is $3\%$. Lower panel (B): The SNR for the all--sky average (results analogous to those of Berti et al.~(2006)) (dashed curve) and the Monte Carlo results with $N=1\, 600$ Kerr sources (solid curve and error bars) at luminosity distance of 3Gpc (redshift of $z=0.522$), for $j=0.8$. Radiation efficiency for the all--sky average case is $4.948\%$.}
\label{sch}
\end{figure}

We first test our Monte Carlo integration for the simple case in which all black holes are assumed to be Schwarzschild black holes ($j=0$).  In particular, we confront our results with the all--sky average results, and study how well they agree. Figure \ref{sch}A shows the SNR for the all--sky average and for the Monte Carlo simulation. 
We find that the two signals overlap, as indeed is expected. In particular, in the Schwarzschild case our Monte Carlo results agree to the expected accuracy level with the results of Berti et al.~(2006).

The Schwarzschild test does not test all sectors of our code, as it does not calculate the spin--weighted spheroidal harmonics. Fixing the value of $j$ we can test the remaining sectors of the code. Figure \ref{sch}B displays the SNR for the all-sky average (for a fixed value of $j$) and the Monte Carlo results, which are in agreement to within the error bars.

\begin{figure}[h]
 \includegraphics[width=3.4in]{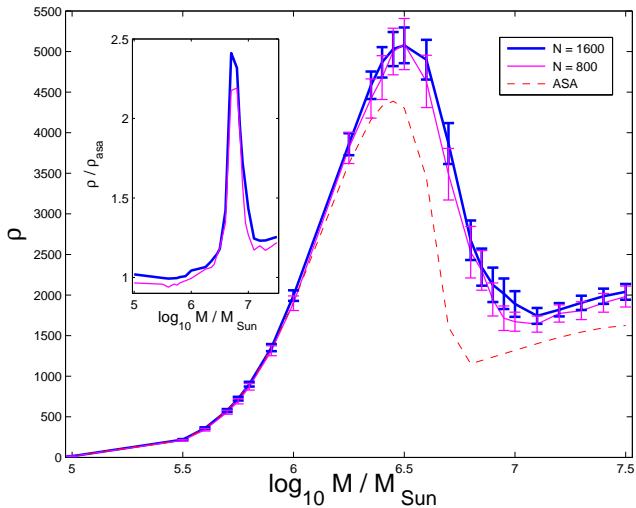} 
\caption{The SNR for the Monte Carlo results with $N$ Kerr sources at variable luminosity distances out to redshift $z=2$ and variable $j$: $N=800$ (thin solid curve) and $N=1\, 600$ (thick solid curve). The corresponding all--sky average results are shown for comparison, at redshift $z=1.509$ (dashed curve). The insert shows the ratio of the Monte Carlo to the all-sky average SNR. }
\label{conv}
\end{figure}

We also test the convergence of our Monte Carlo code, letting all eleven variables be random. Figure \ref{conv} shows the SNR for different Monte Carlo runs, with different sample sizes. The convergence of the result is indeed as expected, namely scaling with $\sqrt{N}$, $N$ being the number of sources in the sample (see a more detailed analysis of  Fig.~\ref{conv} in Section \ref{nasa}).

\section{Non all--sky average results}
\label{nasa}

\begin{figure}[h]
 \includegraphics[width=3.4in]{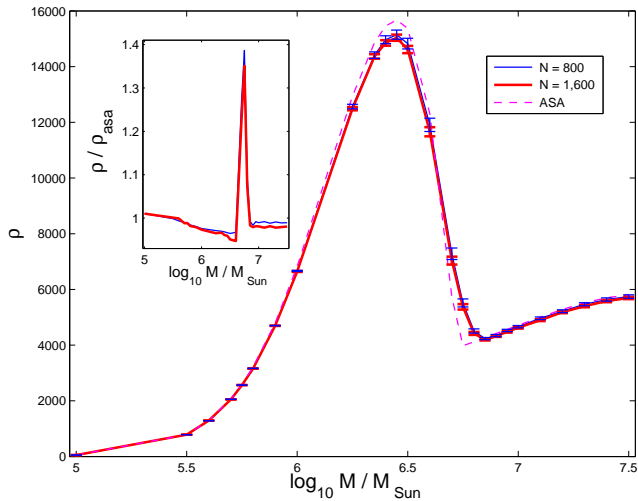} 
\caption{The SNR for the all--sky average (dashed curve) and the Monte Carlo results with variable $j$ for $N$ Kerr sources for fixed parameters other than $j$: $N=800$ (thin solid curve) and $N=1600$ (thick solid curve). 
The insert shows the ratio of the SNRs. The all--sky average results are obtained for $j=0.71882$. Here, all sources are at redshift $z=1.509$, $\cos\iota=\cos\theta=0.7795$, $\beta=\phi=0.1894$, $\psi=0$, $\phi^+=\phi^{\times}=0$, $\epsilon^+=\epsilon^{\times}$, and $\epsilon_{\rm rd}=3.824\%$.}
\label{varj}
\end{figure}

As we already pointed out, we do not expect the behavior of the SNR when $j$ is let to vary to necessarily be identical to the  case of $j$ being averaged. In Figure \ref{varj} we fix all the variables, and let only $j$ be a random variable. The distance of the sources is fixed at redshift of $z=1.509$. Indeed, we find that the behavior is not identical. While for very low and very high masses the behavior is nearly indistinguishable, there is a range of masses, specifically $M\sim 4$--$7\times 10^6M_{\odot}$ for which the SNR in the variable case is significantly higher than in the all--sky average case. The ratio of the two SNRs spikes at $M\sim 5.6\times 10^6M_{\odot}$, where $\rho /\rho_{\rm all-sky}\approx 1.4$ which would translate to an event rate greater by a factor of $\sim 2.75$. This increase in the SNR is $14.2\sigma$, so that it is not an artifact of the Monte Carlo error associated with the finite sample size. 

\begin{figure}[h]
 \includegraphics[width=3.4in]{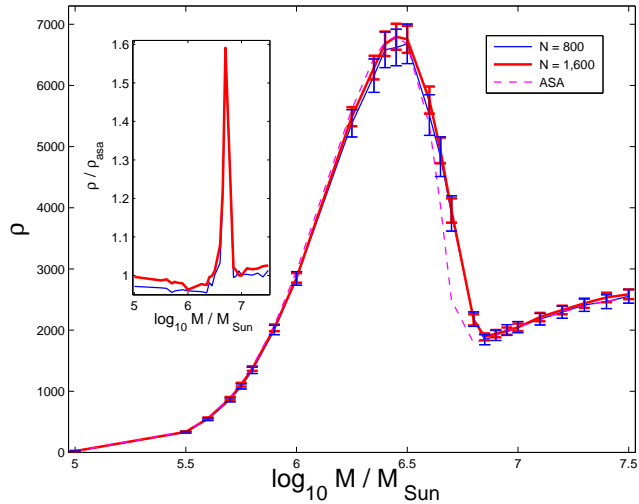} 
\caption{The SNR for the all--sky average  (dashed curve) and the Monte Carlo results with all variables random (except for distance that is fixed at redshift of $z=1.509$) for 800 Kerr sources (thin solid curve and error bars) and for $1\, 600$ Kerr sources (thick solid curve and error bars).  The all--sky average results are obtained for $j=0.71882$.}
\label{fixedr}
\end{figure}

Next, we let all the variables ---except for the distance, which we fix at redshift of $z=1.509$--- be random. Figure \ref{fixedr} shows the SNR, which preserves the behavior we found in Fig.~\ref{varj}. Specifically, the sharp increase in SNR for masses in the range $M\sim 4$--$7\times 10^6M_{\odot}$ is retained. The SNR is increased by a factor of $1.6$, that translates to an increased event rate by a factor of $4.1$. This increase is $7.4\sigma$. Indeed, this is the expected behavior, as all other variables live in independent spaces. Notice, that as the number of dimensions here is higher than in Fig.~\ref{varj}, the errors are commensurately larger. 

We present in Fig.~\ref{conv} the results for all variable being random, including the distance, from the distance to the Virgo cluster ($18.0 \pm 1.2$ Mpc) up to redshift of $z=2$. We find that the SNR is greatly increased compared with all sources being at the average distance (squared) (redshift of $z=1.509$) over a large range of masses, specifically for $M\gtrsim 3.5\times 10^6M_{\odot}$. The increased SNR peaks at $M\sim5\times 10^6M_{\odot}$, with an increase in the SNR by a factor of $2.4$, which translates to an increase in the event rate by a factor of $13.8$. This increase is $9.1\sigma$. As LISA will detect sources at all distances, we believe the increased SNR shown in Fig.~\ref{conv} is closer to the actual SNR for LISA.

\begin{figure}[h]
 \includegraphics[width=3.4in]{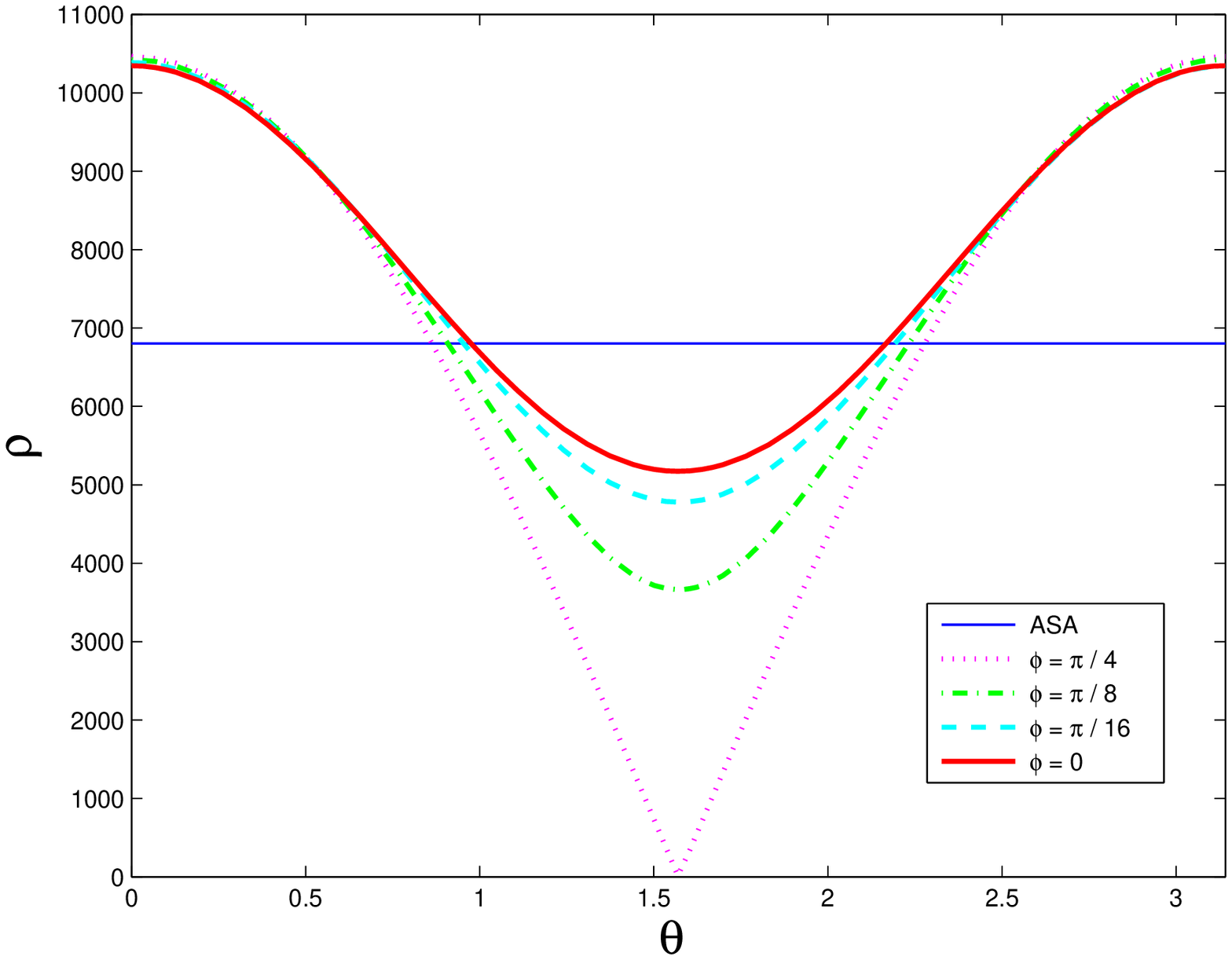} 
\caption{The SNR as a function of the colatitude $\theta$ for different azimuthal directions $\phi$ for $N=800$. Shown are the all--sky average  (thin solid constant line) and the results for signals arriving from specific directions. We show the SNR as a function of $\theta$ for $\phi=0$ (thick solid curve), $\phi=\pi/16$ (dashed curve), $\phi=\pi/8$ (dash-dotted curve), and $\phi=\pi/4$ (dotted curve). 
All data are for Kerr sources with $M=2.8\times 10^6\,M_{\odot}$ at redshift $z=1.509$. All other parameters are obtained from a random sample. The all--sky average results are shown for $j=0.71882$.}
\label{vartheta}
\end{figure}

Most importantly, our approach allows us to find the SNR in different directions in the sky. In Fig.~\ref{vartheta} we show the SNR as a function of the colatitude $\theta$ for various values of $\phi$, along with the all--sky counterpart. We show that the same source (namely, same source's parameters) located in different direction have very different SNR projected on the LISA detector. In particular, there is a LISA blind spot at $\theta=\pi/2$ and $\phi=\pi/4$ (and other values of $\phi$ on the LISA equatorial plane given by the rotational symmetry, namely at $\phi=3\pi/4, 5\pi/4, 7\pi/4$; our numerical results are within $0.1\%$ of these values.). 

This work can be extended in the following ways. First, one may relax the assumption of uniform distribution functions for the random variables, and base the variables on more realistic distribution functions. Second, we use here the delta--function approximation to the quasinormal mode frequency. This assumption may be relaxed (Berti et al.~2006), but the change in the SNR is not expected to be significant, and third, one may extend the single--mode analysis done here to multiple modes (Berti et al.~2006).

The authors are indebted to Vitor Cardoso, Leor Barack, and Emanuele Berti for discussions. This work was supported in part by grants NASA/GSFC NCC5--580 and NASA/SSC NNX07AL52A, by NSF grant No.~PHY--0757344, and by a minigrant from the UAHuntsville Office of the Vice President for Research.

\end{document}